\documentclass[twoside,english]{jfm}
\usepackage[T1]{fontenc}
\usepackage[latin9]{inputenc}
\usepackage[active]{srcltx}
\usepackage{color}
\usepackage{verbatim}
\usepackage{float}
\usepackage{ifsym}
\usepackage{amsmath}
\usepackage{wasysym}
\usepackage{graphicx}
\usepackage{esint}

\makeatletter
\numberwithin{equation}{section}
\numberwithin{figure}{section}

\usepackage{graphicx}
\usepackage{natbib}

\ifCUPmtlplainloaded \else
  \checkfont{eurm10}
  \iffontfound
    \IfFileExists{upmath.sty}
      {\typeout{^^JFound AMS Euler Roman fonts on the system,
                   using the 'upmath' package.^^J}%
       \usepackage{upmath}}
      {\typeout{^^JFound AMS Euler Roman fonts on the system, but you
                   dont seem to have the}%
       \typeout{'upmath' package installed. JFM.cls can take advantage
                 of these fonts,^^Jif you use 'upmath' package.^^J}%
      }
  \else
  \fi
\fi

\ifCUPmtlplainloaded \else
  \checkfont{msam10}
  \iffontfound
    \IfFileExists{amssymb.sty}
      {\typeout{^^JFound AMS Symbol fonts on the system, using the
                'amssymb' package.^^J}%
       \usepackage{amssymb}%
         \let\leq=\leqslant
         
      }{}
  \fi
\fi

\ifCUPmtlplainloaded \else
  \IfFileExists{amsbsy.sty}
    {\typeout{^^JFound the 'amsbsy' package on the system, using it.^^J}%
     \usepackage{amsbsy}}
    {\providecommand\boldsymbol[1]{\mbox{\boldmath $##1$}}}
\fi

\newsavebox{\astrutbox}
\sbox{\astrutbox}{\rule[-5pt]{0pt}{20pt}}

\title[On the kurtosis of water waves]{On the kurtosis of deep-water gravity waves}

\author[Francesco Fedele]%
{Francesco Fedele$^{1,2}$%
  \thanks{Email address for correspondence: fedele@gatech.edu}}

\affiliation{$^1$School of Civil and Environmental Engineering, Georgia Institute of Technology,
Atlanta, GA 30322, USA\\[\affilskip]
$^2$School of Electrical and Computer Engineering, Georgia Institute of Technology, Atlanta, GA 30322, USA\\}

\pubyear{2010}
\volume{650}
\pagerange{119--126}
\date{?; revised ?; accepted ?. - To be entered by editorial office}

\makeatother

\usepackage{babel}
\begin{document}
\maketitle\global\long\def\S{\mathcal{S}}
\global\long\def\eps{\varepsilon}
\global\long\def\H{\mathcal{H}}
\global\long\def\L{\mathcal{L}}
\global\long\def\M{\mathcal{M}}
\global\long\def\K{\mathbf{K}}
\global\long\def\Hilb{\mathbf{H}}
\global\long\def\R{\mathbb{R}}
\global\long\def\ud{\mathrm{d}}

\begin{abstract}
In this paper, we revisit Janssen's (2003) formulation for the dynamic
excess kurtosis of weakly nonlinear gravity waves at deep water. For
narrowband directional spectra, the formulation is given by a sixfold
integral that depends upon the Benjamin-Feir index and the parameter\textcolor{blue}{{}
$R=\sigma_{\theta}^{2}/2\nu^{2}$}, a measure of short-crestedness
for the dominant waves with $\nu${\normalsize{} and $\sigma_{\theta}$}
denoting spectral bandwidth and angular spreading. Our refinement
leads to a new analytical solution for the dynamic kurtosis of narrowband
directional waves described with a Gaussian type spectrum. For multidirectional
or short-crested seas initially homogenous and Gaussian, in a focusing
(defocusing) regime dynamic kurtosis grows initially, attaining a
positive maximum (negative minimum) at the intrinsic time scale
\[
\tau_{c}=\nu^{2}\omega_{0}t_{c}=1/\sqrt{3R},\qquad\mathrm{or}\qquad t_{c}/T_{0}\approx0.13/\nu\sigma_{\theta},
\]
where $\omega_{0}=2\pi/T_{0}$ denotes the dominant angular frequency.
Eventually the dynamic excess kurtosis tends monotonically to zero
as the wave field reaches a quasi-equilibrium state characterized
with nonlinearities mainly due to bound harmonics. Quasi-resonant
interactions are dominant only in unidirectional or long-crested seas
where the longer-time dynamic kurtosis can be larger than that induced
by bound harmonics, especially as the Benjamin-Feir index increases.
Finally, we discuss the implication of these results on the prediction
of rogue waves.\end{abstract}

\section{Introduction}

Third-order quasi-resonant interactions and associated modulational
instabilities cause the statistics of weakly nonlinear gravity waves
to significantly differ from the Gaussian structure of linear seas
(\cite{Janssen2003,Fedele2008a,Onorato2009,ShemerGJR2009,Toffoli2010,XiaoJFM2013}).
One integral statistic used as a measure of the relative importance
of such nonlinearities is the excess kurtosis defined by \cite{Janssen2003}
as 
\begin{equation}
C_{4}=\frac{\left\langle \eta^{4}\right\rangle }{3\sigma^{4}}-1,\label{C40}
\end{equation}
where $\eta$ is the surface displacement with respect to the mean
sea level, $\sigma^{2}=\left\langle \eta^{2}\right\rangle $ is the
wave variance and brackets denote statistical average. In general,
\begin{equation}
C_{4}=C_{4}^{d}+C_{4}^{b},\label{C4total}
\end{equation}
which comprises a dynamic component $C_{4}^{d}$ due to nonlinear
wave-wave interactions (\cite{Janssen2003}) and a bound contribution
$C_{4}^{b}$ induced by the characteristic crest-trough asymmetry
of ocean waves (see e.g. \cite{Tayfun1980,TayfunLo1990,TayfunFedele2007,Fedele2009}).\textcolor{blue}{{}
If third-order Stokes contributions are taken into account (\cite{Janssen2009,Janssen2009JFM,Janssen2014a})
\begin{equation}
C_{4}^{b}=6\mu^{2}.\label{C4b}
\end{equation}
}For unidirectional (long-crested) seas initially homogenous and Gaussian
on deep water, \cite{Janssen2006} have shown that the large-time
behavior of the dynamic excess kurtosis is to monotonically increase
towards the asymptotic value
\begin{equation}
C_{4,NLS}^{d}=BFI^{2}\frac{\pi}{3\sqrt{3}},\label{C4R}
\end{equation}
where
\begin{equation}
BFI=\frac{\mu\sqrt{2}}{\nu}\label{BFI}
\end{equation}
is the Benjamin-Feir index, 
\begin{equation}
\mu=k_{0}\sigma\label{steepness}
\end{equation}
represents an integral measure of wave steepness, $\nu$ is the spectral
bandwidth and $k_{0}$ is the dominant wavenumber. The preceding approximation
is valid for the dynamics of unidirectional narrowband waves described
by one-dimensional (1-D) nonlinear Schrodinger (NLS) and Dysthe (1979)
equations\nocite{Dysthe1979} (see, for example, \cite{ShemerGJR2009,ShemerJGR2010,ShemerPOF2010}).

\textcolor{blue}{Clearly, the preceding results are valid for unidirectional
waves where energy is 'trapped' as in a long wave-guide. If dissipation
is negligible and the wave steepness is small, quasi-resonant interactions
are effective in reshaping the wave spectrum, inducing nonlinear focusing
and large waves in the form of breathers via modulation instability
before breaking occurs (\cite{Onorato2009,ShemerGJR2009,ShemerJGR2010,Chabchoub2011,Chabchoubc2012,ShemerPoF2013,Shemer2014}).
However, such 1-D conditions never occur in nature as they are unrealistic
models of oceanic wind seas. The latter are typically multidirectional
(short-crested) and energy can spread directionally. As a result,
nonlinear focusing due to modulational effects is reduced (\cite{Onorato2009,WasedaJPO2009,Toffoli2010}). }

In regard to the kurtosis in short-crested seas initially homogenous
and Gaussian, the focus of recent numerical studies has been on the
asymptotic behavior with time (see, for example \cite{Shrira2013_JFM,Shrira2014_JPO,Janssen2009}).
Theoretical studies on the transient short-lived features of kurtosis
and their relevance to the prediction of rogue waves are desirable.
These provide the principal motivation for revisiting Janssen's (2003)
formulation for the dynamic excess kurtosis of weakly nonlinear deep-water
gravity waves. 

The remainder of the paper is organized as follows. We first review
Janssen's (2003) dynamic kurtosis model. Then, we present a new analytical
solution of a sixfold integral that yields the growth rate of the
dynamic excess kurtosis for narrowband Gaussian-shaped spectra. This
is followed by a detailed study of its short-time evolution and long-time
asymptotic behavior and comparisons to numerical simulations and experiments.
In concluding, we discuss the implications of these results on rogue
wave prediction.

\section{Dynamic excess kurtosis }

Drawing on \cite{Janssen2003} the dynamic excess kurtosis of weakly
nonlinear sea states, initially homogenous and Gaussian, is given
by
\begin{equation}
C_{4}^{d}=\frac{4g}{\sigma^{2}}\mathrm{Re}\int T_{12}^{34}\delta_{12}^{34}\sqrt{\frac{\omega_{4}}{\omega_{1}\omega_{2}\omega_{3}}}G(t)E_{1}E_{2}E_{3}\mathrm{d}\omega_{1,2,3}\mathrm{d}\theta_{1,2,3},\label{mu4}
\end{equation}
where the resonant function 
\begin{equation}
G(t)=\frac{1-\mathrm{exp}(-i\omega_{12}^{34}t)}{\omega_{12}^{34}},\label{RES}
\end{equation}
$T_{12}^{34}$ is the Zakharov kernel (\cite{Zakharov1968,Zakharov1999,Krasitskii1994})
as a function of the wavenumber vectors $\mathbf{k}_{j}=\left(k_{j}\mathrm{cos}(\theta_{j}),\, k_{j}\mathrm{sin}(\theta_{j})\right)$
and $\mathrm{Re}(x)$ denotes the real part of $x$. The sixfold integral
in Eq. (\ref{mu4}) is defined over the manifold 
\begin{equation}
\mathbf{k}_{1}+\mathbf{k}_{2}-\mathbf{k}_{3}-\mathbf{k}_{4}=\mathbf{0},\label{RESK}
\end{equation}
or equivalently $\delta_{12}^{34}=\delta(\mathbf{k}_{1}+\mathbf{k}_{2}-\mathbf{k}_{3}-\mathbf{k}_{4})$,
where $\delta(\mathbf{k})$ is the Dirac delta. The frequency mismatch
is given by $\omega_{12}^{34}=\omega_{1}+\omega_{2}-\omega_{3}-\omega_{4}$,
$E(\omega,\theta)$ is the surface spectrum and $\sigma^{2}$ is the
variance of surface elevations. The deep-water angular frequency $\omega(k)=\sqrt{gk}$,
the wavenumber magnitude $k=\left|\mathbf{k}\right|$ and 
\[
\omega_{4}=\sqrt{gk_{4}}=\sqrt{g\left|\mathbf{k}_{1}+\mathbf{k}_{2}-\mathbf{k}_{3}\right|}
\]
follows from (\ref{RESK}), with $g$ denoting gravity acceleration.
Since homogenous Gaussian initial conditions with random phases and
amplitudes are imposed, it follows that 
\[
C_{4}^{d}(t=0)=0.
\]
Eq. (\ref{mu4}) can be simplified by resorting to a narrowband approximation
(\cite{Janssen2006,Janssen2009}). So, we assume the spectrum $E$
to peak at $\omega=\omega_{0}$ and $\theta=\theta_{0}$, where $\omega_{0}$
and $\theta_{0}$ denote the dominant angular frequency and wave direction,
respectively, and the associated wavenumber $k_{0}=\omega_{0}^{2}/g$,
wave period $T_{0}=2\pi/\omega_{0}$ and phase speed $c_{0}=\omega_{0}/k_{0}$.
Next, define 
\[
\omega_{j}=\omega_{0}(1+\nu v_{j}),\qquad\theta_{j}=\theta_{0}+\sigma_{\theta}\phi_{j},
\]
where $\nu$ and $\sigma_{\theta}$ denote spectral and angular widths
respectively. Under the narrowband condition $\nu,\sigma_{\theta}\ll1$,
$T_{12}^{34}\sim k_{0}^{3}$ to leading order and the frequency mismatch,
correct to $O(\nu^{2},\sigma_{\theta}^{2})$, is given by
\begin{equation}
\omega_{12}^{34}\sim\nu^{2}\omega_{0}\Delta,\label{wmis}
\end{equation}
with
\[
\Delta=\left\{ (v_{1}-v_{3})(v_{2}-v_{3})-R(\phi_{1}-\phi_{3})(\phi_{2}-\phi_{3})\right\} =\Delta_{v}-R\Delta_{\phi},
\]
where $\Delta_{z}=(z_{1}-z_{3})(z_{2}-z_{3})$ for a generic $z=(z_{1},z_{2},z_{3})$
triplet, and the parameter
\begin{equation}
R=\frac{1}{2}\frac{\sigma_{\theta}^{2}}{\nu^{2}}\label{R}
\end{equation}
is a measure of short-crestedness of dominant waves (\cite{Janssen2009}).
Expanding Eq. (\ref{mu4}) around $\nu=0$ and $\sigma_{\theta}=0$,
to leading order
\begin{equation}
C_{4}^{d}(\tau)=BFI^{2}J(\tau;R),\label{c4}
\end{equation}
where
\begin{equation}
\mathrm{\mathit{J}(\tau;\mathit{R})=2\, Re}\int\frac{1-\mathrm{exp}(i\Delta\tau)}{\Delta}\widetilde{E}_{1}\widetilde{E}_{2}\widetilde{E}_{3}\mathrm{d}v_{1,2,3}\mathrm{d}\phi_{1,2,3}.\label{J}
\end{equation}
Here, $\tau=\nu^{2}\omega_{0}t$ is a dimensionless time and $\widetilde{E}_{j}(v_{j},\phi_{j})=E_{j}/\sigma$. 

For a Gaussian-shaped spectrum, the rate of change of $C_{4}^{d}$
is explicitly given by 
\begin{equation}
\frac{\mathrm{d}C_{4}^{d}}{\mathrm{d}\tau}=BFI^{2}\frac{\mathrm{d}J}{\mathrm{d}\tau},\label{dC4}
\end{equation}
and 
\begin{equation}
\frac{\mathrm{d}J}{\mathrm{d}\tau}=\frac{\mathrm{d}J_{0}(\tau;1,R)}{\mathrm{d}\tau}=2\,\mathrm{Im}\left(\frac{1}{\sqrt{1-2i\tau+3\tau^{2}}\sqrt{1+2iR\tau+3R^{2}\tau^{2}}}\right),\label{dJ}
\end{equation}
where the function $J_{0}(\tau;P,Q)$ is defined in appendix A and
$\mathrm{Im}(x)$ denotes the imaginary part of $x$. \textcolor{blue}{On
this basis, the factor $J$ in Eq. (\ref{c4}) follows by quadrature
as 
\begin{equation}
\mathrm{\mathit{J}(\tau;\mathit{R})=2}\mathrm{\, Im}\int_{0}^{\tau}\frac{1}{\sqrt{1-2i\alpha+3\alpha^{2}}\sqrt{1+2iR\alpha+3R^{2}\alpha^{2}}}d\alpha.\label{JR}
\end{equation}
}

For small times $\tau\ll1$, 
\[
\mathrm{\mathit{J}(\tau;\mathrm{\mathit{R}})=}\int_{0}^{\tau}\left((1-R)\alpha+O(\alpha^{2})\right)d\alpha=\frac{1}{2}(1-R)\tau^{2}
\]
and \textcolor{blue}{Eq. (\ref{c4}) yields}
\begin{equation}
C_{4}^{d}\sim BFI^{2}(1-R)\tau^{2},\quad\quad\tau\ll1,\ \label{smallt}
\end{equation}
in agreement with \cite{Janssen2009}. 

\textcolor{blue}{Note that the dynamic excess kurtosis in Eq. (\ref{c4})
}is consistent with the evolution of weakly nonlinear narrowband wavetrains
of the two-dimensional (2-D) NLS equation.

\section{Intrinsic nonlinear time scale}

The growth rate (\ref{dC4}) of the dynamic $C_{4}^{d}$ vanishes
at the dimensionless time 
\begin{equation}
\tau_{c}=\frac{1}{\sqrt{3R}},\label{tauc}
\end{equation}
or in physical units 
\begin{equation}
\frac{t_{c}}{T_{0}}=\frac{1}{2\pi}\sqrt{\frac{2}{3}}\frac{1}{\sigma_{\theta}\nu}\sim\frac{0.13}{\sigma_{\theta}\nu},\label{tc}
\end{equation}
where $T_{0}=2\pi/\omega_{0}$ is the dominant wave period. Further,
the second derivative of $C_{4}^{d}$ at $\tau_{c}$ is given by 
\[
\left.\frac{\mathrm{d}^{2}C_{4}^{d}}{\mathrm{d}\tau^{2}}\right|_{\tau=\tau_{c}}=-6\sqrt{3}R(1-R).
\]
Thus, $C_{4}^{d}$ attains a positive maximum (negative minimum) at
$\tau=\tau_{c}$ for $0<R<1$ ($R>1$). It is straightforward to show
that for multidirectional or short-crested seas ($R>0$) 
\[
\underset{\tau\rightarrow\infty}{\mathrm{lim}}C_{4}^{d}=0.
\]
Indeed, it is sufficient to study the rate of change of $C_{4}^{d}$
for large times $\tau\gg1$. To do so, consider the change of variable
$\tau=1/r$ and expanding (\ref{dC4}) around $r=0$ yields

\begin{equation}
\frac{\mathrm{d}C_{4}^{d}}{\mathrm{d}\tau}\sim\frac{(-1+R)r^{3}}{9R^{2}}=\frac{(-1+R)}{9R^{2}\tau^{3}}.\label{rate}
\end{equation}
Note that the real part of Eq. (\ref{dC4}), which has no physical
meaning, decays as $\tau^{-2}$. For $0<R<1$, $C_{4}^{d}$ first
attains a positive peak at $\tau=\tau_{c}$ and then decays monotonically
to zero since $\mathrm{d}C_{4}^{d}/\mathrm{d}\tau<0$ for large $\tau$.
This is clearly seen in Figure (3.1), showing the evolution of $C_{4}^{d}$
for different values of $R$. For $R>1$, $C_{4}^{d}$ initially decreases
reaching a negative peak at $\tau=\tau_{c}$ and then tends monotonically
to zero, because $\mathrm{d}C_{4}^{d}/\mathrm{d}\tau>0$ for large
$\tau$ as shown in Figure (3.1). At the critical value $R=1$, the
excess kurtosis is null at any time, as can easily be verified from
Eq. (\ref{dC4}). 

In summary, depending on the value of $R$ there will be nonlinear
focussing ($C_{4}^{d}>0$) or nonlinear defocussing ($C_{4}^{d}<0$)
in agreement with \cite{Janssen2009}. Note that for unidirectional
or long-crested seas ($R=0$) the rate of change $\mathrm{d}C_{4}^{d}/\mathrm{d}\tau>0$
for any time $\tau$. In this case, the dynamic excess kurtosis monotonically
increases with time to the asymptotic value of Eq. (\ref{C4R}) (\cite{Janssen2006,ShemerGJR2009,ShemerJGR2010,ShemerPOF2010,fedeleNLS}). 

\begin{figure}[H]
\centering\includegraphics[scale=0.45]{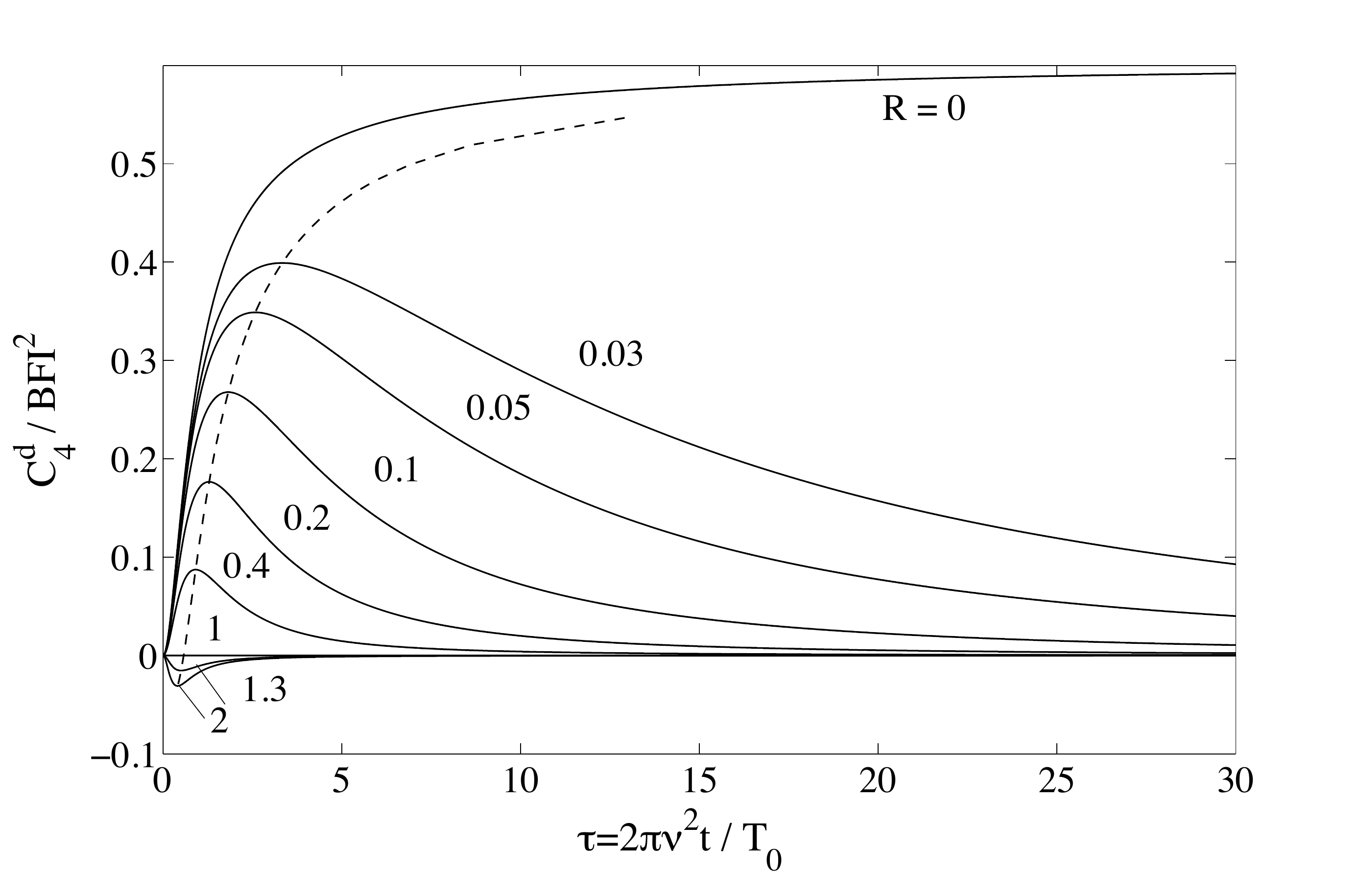} \protect\caption{\textcolor{blue}{Dynamic excess kurtosis: (solid lines) $C_{4}^{d}/BFI^{2}$
as a function of dimensionless time $\tau=\nu^{2}\omega_{0}t=2\pi\nu^{2}t/T_{0}$
for different values of $R$; (dashed line) locus of transient peaks
($T_{0}$ denotes the dominant wave period and $\nu$ is the spectral
bandwidth). }}

\label{FIG1} 
\end{figure}

\section{Dynamic excess kurtosis maximum}

From Eqs. (\ref{c4}) and (\ref{tauc}), the peak value of $C_{4}^{d}$
at $\tau=\tau_{c}$ is given by

\begin{equation}
C_{4}^{\mathrm{\mathit{d}}}(R)=BFI^{2}J_{p}(R),\label{Cmax}
\end{equation}
where

\[
\mathrm{\mathit{J}_{\mathit{p}}(\mathrm{\mathit{R}})=\mathit{\mathrm{\mathit{J}\left(\frac{1}{\sqrt{3\mathit{R}}};\mathit{R}\right)}}=Im}\int_{0}^{\frac{1}{\sqrt{3R}}}\frac{2}{\sqrt{1-2i\alpha+3\alpha^{2}}\sqrt{1+2iR\alpha+3R^{2}\alpha^{2}}}d\alpha.
\]
The following relation holds
\[
J_{\mathrm{\mathit{p}}}\left(\frac{1}{R}\right)=-RJ_{\mathrm{\mathit{p}}}(R),
\]
in agreement with \cite{Janssen2009}. This relation allows us to
compute the minimum kurtosis for $R>1$ from the maximum value for
$R<1$. Indeed, 

\begin{equation}
C_{4,\mathrm{min}}^{\mathrm{\mathit{d}}}\left(\frac{1}{R}\right)=-RC_{4,\mathrm{max}}^{\mathrm{\mathit{d}}}(R),\qquad\qquad0\leq R\leq1.\label{minC}
\end{equation}
where $C_{4,\mathrm{max}}^{\mathrm{\mathit{d}}}=BFI^{2}J_{p}(R)$.
Clearly, this vanishes at $R=1$ signaling the change from a nonlinear
focusing to defocusing regime where the dynamic kurtosis is negative. 

Drawing on \cite{Janssen2009}, the limit 

\begin{equation}
\mathrm{\mathit{J_{\mathrm{\mathit{p}}}(R)}}\sim-\frac{\pi}{3\sqrt{3}R},\qquad\qquad R\gg1\label{Rinf}
\end{equation}
and that for small-times in Eq. (\ref{smallt}) suggest the least-squares
fit for the maximum 

\begin{equation}
\frac{C_{4,\mathrm{max}}^{\mathrm{\mathit{d}}}(R)}{BFI^{2}}=J_{\mathrm{peak}}(R)\approx\frac{b}{(2\pi)^{2}}\frac{1-R}{R+bR_{0}},\qquad\qquad0\leq R\leq1,\label{fit}
\end{equation}
where $R_{0}=\frac{3\sqrt{3}}{\pi}$ and $b=2.48$. In the left panel
of Figure 4.1, the preceding approximation is compared against the
theoretical $C_{4,\mathrm{max}}^{\mathrm{\mathit{d}}}$ solving Eq.
(\ref{Cmax}) by numerical integration. Evidently, the latter is slightly
larger than the maximum excess kurtosis derived by \cite{Janssen2009},
who have also used (\ref{fit}) but with $b=1$. Their maximum follows
by first taking the limit of the resonant function $G(t)$ in Eq.
(\ref{RES}) at $t=\infty$ and then solving the sixfold integral
in Eq. (\ref{mu4}). Clearly, for $R>0$ the dynamic excess kurtosis
should vanish at large times as discussed above.\textcolor{blue}{}%
\textcolor{blue}{{} }Janssen (personal communication, 2014\nocite{Janssen2014})
confirmed that Eq. (\ref{Cmax}) holds and provided an alternative
proof that $C_{4}^{d}$ tends to zero as $t\rightarrow\infty$ using
complex analysis and numerical integration.

Further, from (\ref{tauc})
\begin{equation}
\frac{C_{4,\mathrm{max}}^{\mathrm{\mathit{d}}}(\tau_{c})}{BFI^{2}}\approx\frac{b}{(2\pi)^{2}}\frac{-1+3\tau_{c}^{2}}{1+3bR_{0}\tau_{c}^{2}},\qquad\qquad0\leq\tau_{c}\leq\frac{1}{\sqrt{3}}.\label{fit2}
\end{equation}
Clearly, the transient maximum kurtosis becomes larger for longer
time scales $\tau_{c}$, as illustrated in the right panel of Figure
4.1. Note that the dynamic excess kurtosis is negative for $\tau_{c}>1/\sqrt{3}$
as the wave regime is of defocusing type ($R>1$) and the minimum
value $C_{4,\mathrm{min}}^{\mathrm{\mathit{d}}}$ can be computed
from Eq. (\ref{minC}).

\begin{figure}[H]
\centering\includegraphics[width=1\textwidth]{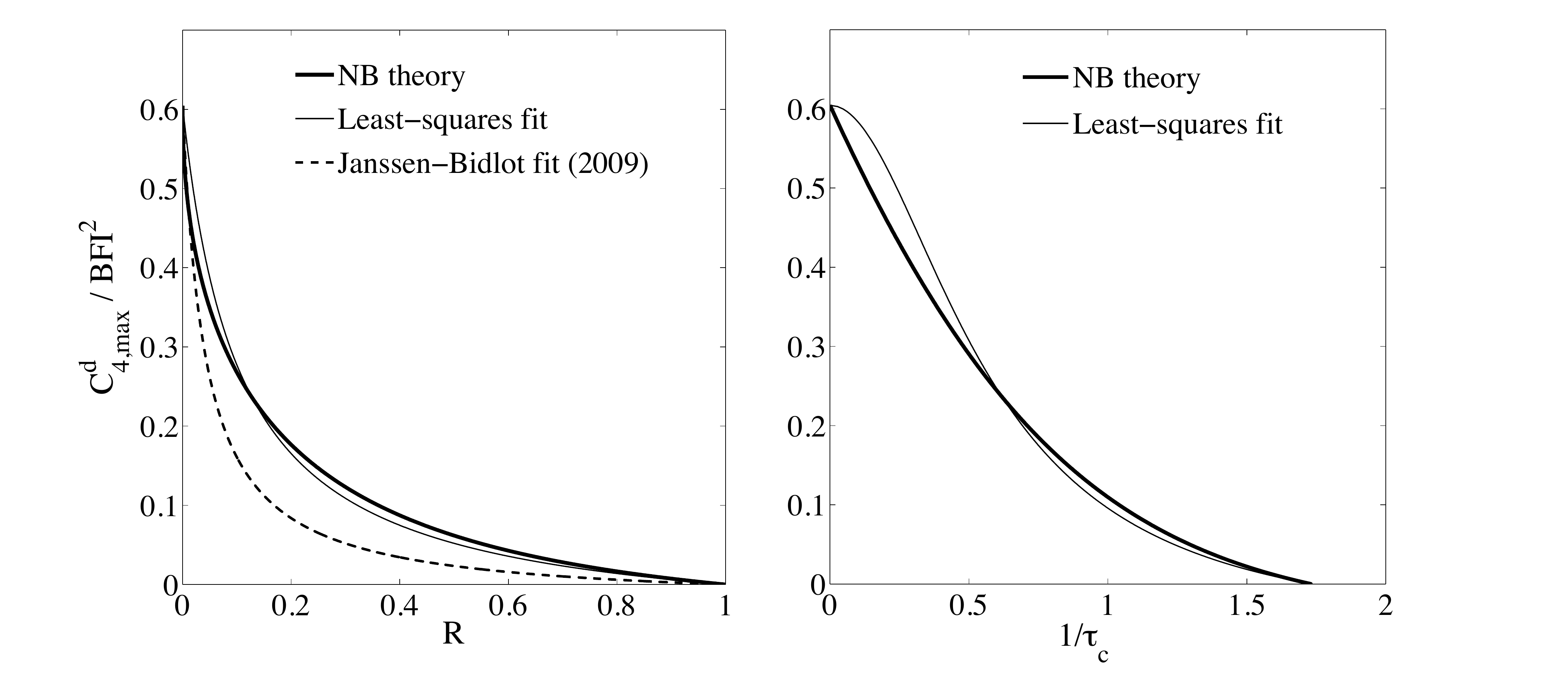} \protect\caption{Maximum dynamic excess kurtosis $C_{4,\mathrm{max}}^{\mathrm{\mathit{d}}}$
as a function of (left) $R$ and (right) $1/\tau_{c}$: (bold line)
present theoretical prediction, (thin line) least-squares fit from
Eq. (\ref{fit}) ($b=2.48$) and (dash line) Janssen-Bidlot (2009)
fit ($b=1$). }

\label{FIG3a} 
\end{figure}

\section{Comparisons to simulations and experiments}

We now compare the theoretical narrowband (NB) predictions for the
total kurtosis $C_{4}$ {[}see Eqs. (\ref{C4total}),(\ref{c4}) and
(\ref{C4b}){]} to experimental results (\cite{Onorato2009}) and
the comprehensive numerical simulations of JONSWAP directional wave
fields carried out by \cite{XiaoJFM2013} and \cite{Toffoli2010}.
They considered the broad-band modified nonlinear Schrodinger equations
(BMNLS) (\cite{Dysthe1979}) and a high-order spectral (HOS) solver
(\cite{DommermuthYue1987HOS}). In particular, we consider the comprehensive
numerical results reported in Figs. 10a,b in \cite{XiaoJFM2013} for
the two cases of narrow and broad directional spreading, i.e. $\sigma_{\theta}=0.04$
and 0.07 respectively. The simulated sea states have standard deviation
$\sigma=0.02$ m, dominant wave period $T_{0}=1$ s, significant wave
height $H_{s}=4\sigma=0.08$ m, $BFI=0.78$, wave steepness $\mu=0.08$
and spectral bandwidth $\nu=0.15$ (see appendix B for the estimation
of wave parameters). As shown in Figure 5.1, the numerical studies
by \cite{XiaoJFM2013} indicate an initial overshoot of the kurtosis
followed by a decay towards quasi-Gaussian conditions. 

In particular, the left panel of Figure 5.1 shows that for a narrow
directional spreading ($\sigma_{\theta}\sim0.04$) the present theoretical
NB model (thick line) explains the peak kurtosis and the initial transient
behavior of BMNLS simulations (thin dashed line) as NB is consistent
with the dynamics of the 2-D NLS equation. BMNLS and NB yield faster
initial growth and overestimate both HOS (thin solid line) and experiments
(triangle symbols). However, soon after the transient stage, the spectrum
has already broadened in frequency and spread angularly, approaching
a quasi-equilibrium state. At this stage, the NB approximation provides
just a qualitative trend of the large-time behavior since it does
not account for spectral changes. In particular, NB shows a slower
decaying trend to zero than BMNLS. This indicates that numerical models
capture the directional energy spreading and quasi-resonant interactions
attenuate much faster than NB after the transient peak. 

For a broad directional spreading ($\sigma_{\theta}\sim0.07$) the
right panel of Figure 5.1 shows that NB overestimates the maximum
kurtosis and qualitatively explains the initial transient overshoot
of BMNLS simulations, which are now beyond their range of validity
as the spectrum is already too broad initially. Instead, HOS simulations
are in agreement with experiments and yield a smaller value of the
maximum kurtosis and a slower transient than BMNLS. 

In both the abovementioned cases, the NB model qualitatively describes
the initial transient and kurtosis peak. For time scales $t\gg t_{c}$,
NB indicates the correct asymptotic behavior of the total kurtosis
of surface elevations as dominated by nonlinear bound harmonics (see
also \cite{Shrira2013_JFM,Shrira2014_JPO}).%
{} 

\begin{figure}[H]
\centering\includegraphics[width=1\textwidth]{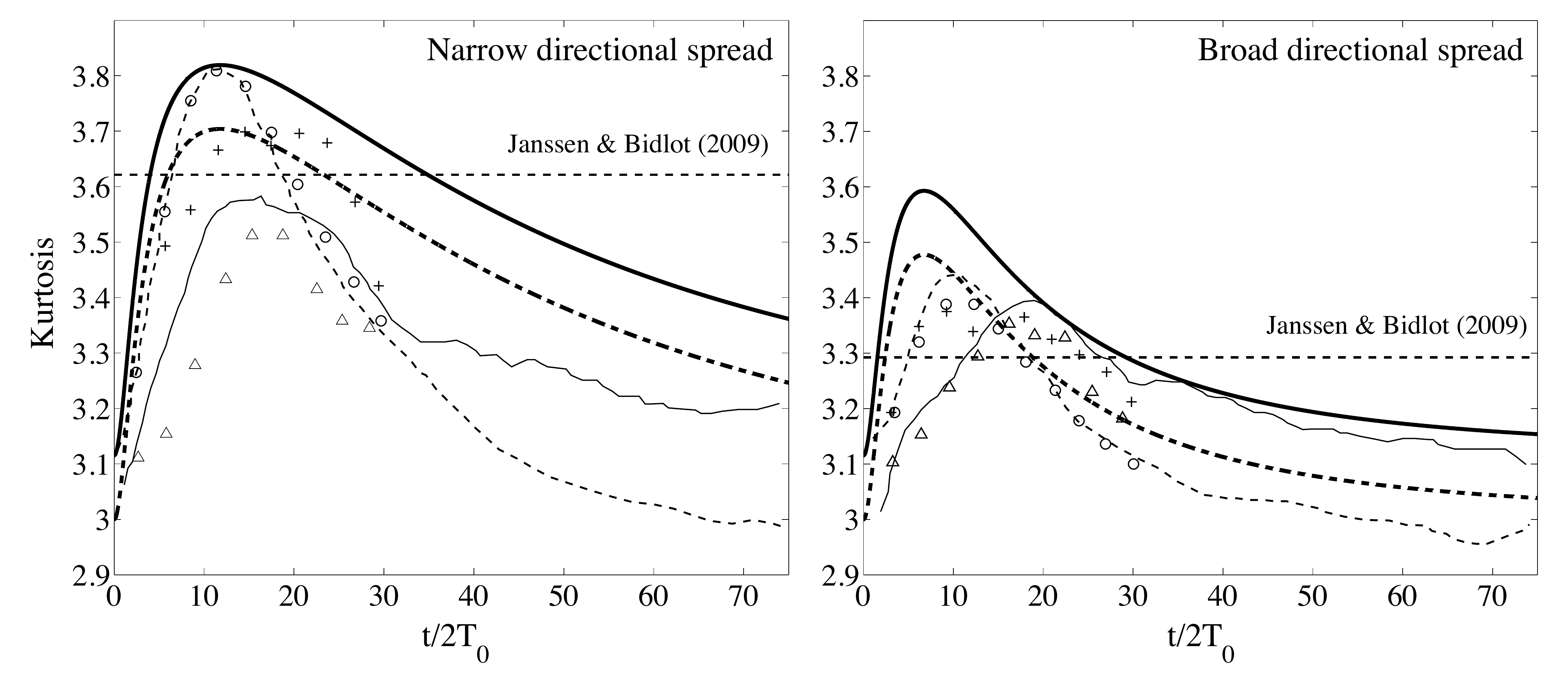} \protect\caption{Kurtosis $\mu_{4}=\left\langle \eta^{4}\right\rangle /\left\langle \eta^{2}\right\rangle ^{2}=3(C_{4}+1)$
as a function of time $t/2T_{0}$ for JONSWAP directional wave fields
initially homogenous and Gaussian ($BFI=0.78$, $\mu=0.08$, $\nu=0.15$):
theoretical narrowband predictions compared against simulations and
experiments (\textifsymbol[ifgeo]{49}) from \cite{Onorato2009} (data
digitized from Fig. 10a,b in \cite{XiaoJFM2013}). (Left) narrow directional
spreading with $\sigma_{\theta}=0.04$, $R=0.03$ (see Eq. (\ref{R}))
and (Right) broad directional spreading with $\sigma_{\theta}=0.07$,
$R=0.1$. Narrowband theory: dynamic kurtosis $\mu_{4}^{d}=3(C_{4}^{d}+1)$
from Eq. (\ref{c4}) (thick dashed line) and total kurtosis $\mu_{4}=\mu_{4}^{d}$+$\mu_{4}^{b}$
(thick solid line), with $\mu_{4}^{b}=3(C_{4}^{b}+1)$ from Eq. (\ref{C4b}).
\textcolor{blue}{Dashed horizontal lines denote Janssen \& Bidlot
(2009) dynamic kurtosis maximum from Eq. (\ref{fit}) with $b=1$.}
Simulation results from \cite{XiaoJFM2013}: HOS (thin solid line),
BMNLS (thin dashed line). The numerical results from \cite{Toffoli2010}
are also shown: BMNLS (\Circle ) and HOS (+). }

\label{FIG2} 
\end{figure}

\section{Concluding remarks}

Our refinement of Janssen's (2003) formulation leads to a new analytical
solution for the dynamic excess kurtosis of weakly nonlinear deep-water
gravity waves and associated growth rate. It assumes that waves are
approximately narrowband and characterized with a Gaussian type spectrum
that does not change over time.

For short-crested seas initially homogenous and Gaussian, in the focusing
regime ($0<R<1$) the dynamic excess kurtosis initially grows attaining
a maximum at the intrinsic time scale 
\begin{equation}
\tau_{c}=\nu^{2}\omega_{0}t_{c}=\frac{1}{\sqrt{3R}},\qquad\mathrm{or}\qquad\frac{t_{c}}{T_{0}}\sim\frac{0.13}{\nu\sigma_{\theta}}.\label{tau}
\end{equation}
Eventually it tends monotonically to zero as energy spreads directionally.
In the defocusing regime ($R>1$) the dynamic excess kurtosis is always
negative attaining a minimum at $t_{c}$ and then it tends to zero
in the long time. Thus, the present theoretical narrowband predictions
indicate a decaying trend for the dynamic excess kurtosis over large
times. This implies that for time scales $t\gg t_{c}$ the asymptotic
behavior of the total kurtosis of surface elevations is dominated
by nonlinear bound harmonics in qualitatively accord with numerical
simulations (\cite{ShriraGRL2009,Toffoli2010,XiaoJFM2013,Shrira2014_JPO})
and experiments (\cite{Onorato2009,WasedaJPO2009}). For time scales
of the order of or less than $t_{c}$ the dynamic component can dominate
and the wave field may experience rogue wave behavior induced by quasi-resonant
interactions (\cite{Janssen2003}). %

\textcolor{blue}{Current statistical approaches for freak wave warning
systems and predictions rely on the Gram-Charlier type probability
distribution for crest-to-trough wave heights $z=H/\sigma$ ( see,
e.g. \cite{Janssen2003,Janssen2006,TayfunFedele2007})
\begin{equation}
\mathrm{Pr}\{H/\sigma>z\}=\left[1+\frac{C_{4}}{384}z^{2}\left(z^{2}-16\right)\right]\mathrm{exp}\left(-\frac{z^{2}}{8}\right)\label{pdf-1}
\end{equation}
and Janssen's (2003) theory for the kurtosis $C_{4}$, a key result
with significant implications to the understanding of the role of
nonlinear wave interactions (\cite{Janssen2003,Janssen2009JFM}).
The present study suggests that it is important to reconsider it carefully.
Indeed, the large excess kurtosis transient observed during the initial
stage is a result of the unrealistic assumption that the initial wave
field is homogeneous Gaussian. A random wave field forgets its initial
conditions and adjusts to a non-Gaussian state dominated by bound
nonlinearities on time scales $t\gg t_{c}$ in agreement with experiments
(\cite{Onorato2009,WasedaJPO2009}) and simulations (\cite{Shrira2013_JFM,Shrira2014_JPO}).
In this regime, statistical prediction of extreme waves can be based
on the Tayfun (1980) model (\cite{TayfunFedele2007,Fedele2008a,Fedele2009,Fedele2015}). }

\textcolor{blue}{The NB approximation for kurtosis is only consistent
with the dynamics of the 2-D NLS equation. It just qualitatively captures
the transient behavior of the kurtosis observed in wave tank experiments
due to a cold start, i.e. initially when the wave field is homogenous
and Gaussian (\cite{Onorato2009}). }Instead, it tends to agree with
numerical simulations of the BMNLS equations for narrowband spectra
(\cite{Toffoli2010,XiaoJFM2013}). And, for time scales $t>t_{c}$
NB theory indicates the correct asymptotic behavior of the total kurtosis
of surface elevations as dominated by bound harmonic contributions. 

\textcolor{blue}{Further, NB predictions tend to overestimate the
observed kurtosis maximum in wave tank experiments (see Fig. (\ref{FIG2}))
suggesting that higher order nonlinearities and broader spectral bandwidth
effects should be accounted for in the theoretical analysis. Indeed,
for }the compact form of the 1-D Zakharov equation (cDZ, \cite{Dyachenko2011}),
\cite{Fedele2014} showed that, correct to $O(\nu^{2})$ in spectral
bandwidth, the associated maximum dynamic kurtosis 
\begin{equation}
C_{4,cDZ}^{d}=C_{4,NLS}^{d}\left(1-\frac{4\sqrt{3}+\pi}{8\pi}\nu^{2}\right)\approx C_{4,NLS}^{d}\left(1-0.40\nu^{2}\right)\label{CDZ-1}
\end{equation}
is smaller than the NLS counterpart $C_{4,NLS}^{d}$ in Eq. (\ref{C4R}),
especially as the spectrum widens.\textcolor{blue}{{} The present study
can be extended to derive an analytical solution of the kurtosis evolution
from a cold start in accord with the 2-D Zakharov equation (\cite{Dyachenko2011,Gramstad2014}),
but this is beyond the scope of this work. }

Thus, the present theoretical results for the third-order nonlinear
statistics of wave fields characterized by a narrow spectrum of Gaussian
shape \textcolor{blue}{are not relevant} for predictions of extreme
waves in realistic oceanic seas. It appears that such results may
just provide a qualitative trend of the short-time kurtosis behavior
induced by a cold start \textcolor{blue}{in the context of experiments
in wave tanks.}

\section{Acknowledgments}

FF is grateful to Peter A. E. M. Janssen for suggesting the topic
of this work and for discussions on nonlinear water waves. FF also
thanks Michael Banner, Victor Shrira and M. Aziz Tayfun for discussions
on nonlinear wave statistics and random wave fields.

\section{Appendix A}

Consider the generic sixfold integral 
\begin{equation}
\mathrm{\mathit{J_{\mathrm{0}}}(\tau;\mathit{P,Q})=2\, Re}\int\frac{1-\mathrm{exp}(i\Delta\tau)}{\Delta}\widetilde{E}_{1}\widetilde{E}_{2}\widetilde{E}_{3}\mathrm{d}v_{1,2,3}\mathrm{d}\phi_{1,2,3}.\label{J-1}
\end{equation}
where $P$ and $Q$ are complex coefficients, 
\[
\Delta=P\Delta_{v}-Q\Delta_{\phi},
\]
and
\[
\widetilde{E}_{j}(v_{j},\phi_{j})=\frac{\mathrm{exp}\left(-\frac{v_{j}^{2}+\phi_{j}^{2}}{2}\right)}{2\pi}.
\]
Then, the integral (\ref{J-1}) can be written as 
\begin{equation}
J_{0}(\tau;P,Q)=\mathrm{Re}\int\frac{1-\mathrm{exp}(iP\Delta_{v}\tau-iQ\Delta_{\phi}\tau)}{P\Delta_{v}-iQ\Delta_{\phi}}\frac{\mathrm{exp}\left(-\frac{v_{1}^{2}+v_{2}^{2}+v_{3}^{2}}{2}\right)}{\left(2\pi\right)^{3/2}}\frac{\mathrm{exp}\left(-\frac{\phi_{1}^{2}+\phi_{2}^{2}+\phi_{3}^{2}}{2}\right)}{\left(2\pi\right)^{3/2}}dv_{1,2,3}d\phi_{1,2,3}\label{c4-1-1}
\end{equation}
Clearly, $v_{j}$ and $\phi_{j}$ are coupled via the denominator
$P\Delta_{v}-iQ\Delta_{\phi}$. However, they become uncoupled if
we take the time derivative

\[
\frac{dJ_{0}}{d\tau}=\mathrm{Im}\int\mathrm{exp}(iP\Delta_{v}\tau-iQ\Delta\tau)\frac{\mathrm{exp}\left(-\frac{v_{1}^{2}+v_{2}^{2}+v_{3}^{2}}{2}\right)}{\left(2\pi\right)^{3/2}}\frac{\mathrm{exp}\left(-\frac{\phi_{1}^{2}+\phi_{2}^{2}+\phi_{3}^{2}}{2}\right)}{\left(2\pi\right)^{3/2}}dv_{1,2,3}d\phi_{1,2,3}
\]
Indeed, 
\begin{equation}
\frac{dJ_{0}}{d\tau}=2\,\mathrm{Im}\left[I_{0}(\tau;P)I_{0}(\tau;-Q)\right]\label{dj}
\end{equation}
where
\[
I_{0}(\tau;P)=\int\mathrm{exp}(iP\Delta_{z}\tau)\frac{\mathrm{exp}\left(-\frac{z_{1}^{2}+z_{2}^{2}+z_{3}^{2}}{2}\right)}{\left(2\pi\right)^{3/2}}dz_{1,2,3}.
\]
Drawing on \cite{fedeleNLS} Gaussian integration yields 
\[
I_{0}(\tau;P)=\frac{1}{\sqrt{1-2iP\tau+3P^{2}\tau^{2}}}
\]
and from (\ref{dj})

\[
\frac{\mathrm{d}J_{0}(\tau;P,Q)}{\mathrm{d}\tau}=\mathrm{2\, Im}\left(\frac{1}{\sqrt{1-2iP\tau+3P^{2}\tau^{2}}\sqrt{1+2iQ\tau+3Q^{2}\tau^{2}}}\right).
\]

\section{Appendix B}

Recently, \cite{XiaoJFM2013} and \cite{Toffoli2010} have compared
BMNLS and HOS simulations of JONSWAP directional wave fields to the
experimental results in \cite{Onorato2009}. Their Benjamin-Feir index
is a factor $\sqrt{2}$ larger than the one used in this work {[}see
Eq. (\ref{BFI}){]}, that is 
\[
BFI'=\frac{2k_{0}\sigma}{\nu}=\frac{2\mu}{\nu}=\sqrt{2}BFI.
\]
Further, their wave steepness $\mu'=2\mu$ where $\mu=k_{0}\sigma$
is used in this work (see also Table 1 in \cite{Toffoli2010}). In
the numerical results reported in Fig. 10a,b of \cite{XiaoJFM2013},
$BFI'=1.1$ and $\mu'=0.16$. Thus, $BFI=0.78$, $\mu=0.08$ and the
spectral bandwidth follows as $\nu=\sqrt{2}\mu/BFI=0.15$. 

The directional distribution $D(\theta)$ adopted by \cite{XiaoJFM2013}
is given by
\[
D(\theta)=\frac{2}{\Theta}\mathrm{cos^{2}}\left(\frac{\pi\theta}{\Theta}\right),\qquad\left|\theta\right|\leq\frac{\Theta}{2},
\]
and the associated directional spreading follows as
\begin{equation}
\sigma_{\theta}=\sqrt{\frac{\int_{-\frac{\Theta}{2}}^{\frac{\Theta}{2}}D(\theta)\theta^{2}d\theta}{\int_{-\frac{\Theta}{2}}^{\frac{\Theta}{2}}D(\theta)d\theta}}=\Theta\sqrt{\frac{\pi^{2}-6}{12\pi^{2}}}.\label{th}
\end{equation}
The numerical results shown in Fig. 10a of \cite{XiaoJFM2013} are
for $\Theta=12\frac{\pi}{180}$ rad (narrow directional spreading);
using Eq. (\ref{th}) yields $\sigma_{\theta}=0.04$ and $R=0.03$
from Eq. (\ref{R}). For the case of broad directional spreading shown
in their Fig. 10b $\Theta=21\frac{\pi}{180}$ rad and $\sigma_{\theta}=0.07$,
$R=0.1$.

\bibliographystyle{jfm}
\bibliography{biblioFranco}

\end{document}